\documentclass[11pt,a4paper]{article}

\usepackage{latexsym}
\usepackage{amssymb}
\usepackage{amsmath}
\usepackage{amsfonts}
\usepackage{mathrsfs}
\usepackage{cite}

\newcommand{\be}{\begin{equation}}
\newcommand{\ee}{\end{equation}}

\def\a{\alpha}
\def\b{\beta}

\def\t{\theta}
\def\tr{{\rm tr}}
\def\tr{{\rm tr}\,}
\def\Tr{{\rm Tr}\,}
\def\cN{{\cal N}}

\def\bea{\begin{eqnarray}}
\def\eea{\end{eqnarray}}
\def\nn{\nonumber}

\def\cN{{\cal N}}

\def\f{\frac}

\def\tr{{\rm tr}\,}

\def\nn{\nonumber}

\def\d{\delta}
\def\q{\quad}
\def\g{\gamma}

\def\U{\Upsilon}
\def\sB{\stackrel{\frown}{\square}}

\def\r{\rho}

\def\eq{\eqref}
\def\pr{\partial}
\def\nb{\nabla}

 \numberwithin{equation}{section}

\begin{document}
\begin{titlepage}

\begin{center}
\vspace{1cm} {\Large\bf Low-energy $6D$, $\cN=(1,1)$ SYM effective
action \\
\vspace{0.2cm}

beyond the leading approximation} \vspace{1.5cm}

 {\bf
 I.L. Buchbinder\footnote{joseph@tspu.edu.ru }$^{\,a,b}$,
 E.A. Ivanov\footnote{eivanov@theor.jinr.ru}$^{\,c}$,
 B.S. Merzlikin\footnote{merzlikin@tspu.edu.ru}$^{\,d,a}$
 }
\vspace{0.4cm}

{\it $^a$ Department of Theoretical Physics, Tomsk State Pedagogical
University,\\ 634061, Tomsk,  Russia \\
 \vskip 0.1cm
 $^b$ National Research Tomsk State University, 634050, Tomsk, Russia \\
 \vskip 0.15cm
 $^c$ Bogoliubov Laboratory of Theoretical Physics, JINR, 141980 Dubna, Moscow region,
 Russia \\ \vskip 0.15cm
 $^d$ Tomsk State University of Control Systems and Radioelectronics, 634050 Tomsk, Russia
 }
\end{center}
\vspace{0.4cm}

\begin{abstract}
For $6D$, $\cN=(1,1)$ SYM theory formulated in $\cN=(1,0)$ harmonic
superspace as a theory of interacting gauge multiplet and
hypermultiplet we construct the $\cN=(1,1)$ supersymmetric
Heisenberg-Euler-type superfield effective action.  The effective
action is computed  for the slowly varying on-shell background
fields and involves,  in the bosonic sector, all powers of a
constant abelian strength.
\end{abstract}

\end{titlepage}

\setcounter{footnote}{0} \setcounter{page}{1}


\section{Introduction}

The low-energy effective action in quantum field theory is a
functional of the slowly varying strengths of the vector gauge
fields and the matter fields (see e.g., \cite{W}). It provides
quantum corrections to the ``microscopic'' action of  given model.
The effective action can serve as a bridge between superstring
theory and supersymmetric gauge theory. On the one hand, some kind
of effective action can be evaluated in string theory; on the other
hand, it can be calculated in the framework of the field theory. A a
result, there emerges a principal possibility to study the
low-energy effects in string theory by field theory methods (see
\cite{Tseytlin} for review of the effective action in string
theory).

The first example of effective action for a quantum field in a
constant external electromagnetic field was constructed by
Heisenberg and Euler in the pioneer paper \cite{HE}. Later, it was
reformulated in a covariant way by Schwinger \cite{Sch}. The
Heisenberg-Euler effective action provides the quantum corrections
to the Maxwell equations involving all powers of the field strength.
The remarkable feature of the Heisenberg-Euler effective action is
its non-perturbative dependence on the coupling constant. Later on,
the Heisenberg-Euler effective action was computed in one-and
two-loop approximations in different field theory models and
employed  for studying their various properties, such as quantum
corrections to the classical equations of motion, particle creation
in external fields, finding the low-energy amplitudes, {\it etc}  (
see,
 e.g., \cite{Dune} for a general review, and \cite{BIP-rev}, \cite{BISam}
 for a review of some supersymmetric applications).

In our previous work \cite{BIM} we have computed the leading
low-energy contribution to the one-loop effective action of  the
six-dimensional $\cN=(1,1)$ SYM theory in the harmonic superspace
approach\footnote{These results were recently confirmed by the
component calculations in  \cite{BBM}.}. The contributions to
effective action we have found include, in the bosonic sector, the
leading terms of the fourth order in the abelian gauge field
strength $F_{MN}$. Such an effective action is in a correspondence
with the massless gluon amplitudes in  $6D, \, \cN=(1,1)$ SYM theory
and is related to the tree-level amplitudes of the massless string
modes in the double-scaled little string theory \cite{Yin1},
\cite{Yin2} (see \cite{Ah}, \cite{Kut} for a review of little
strings)\footnote{The relationships of the $6D, \cN=(1,1)$ SYM
theory with the  low-energy dynamics of D5 branes are discussed in
\cite{L}, \cite{BL}, \cite{GK}.}.

The present paper is a natural continuation  of \cite{BIM}. We
calculate the total superfield Heisenberg-Euler-type effective
action for $\cN=(1,1)$ SYM theory. This effective action can
hopefully be relevant to the little string theory and admit an
equivalent formulation within its context.

Like in our previous publications, we deal with  the six-dimensional
$\cN=(1,1)$ supersymmetric gauge theory formulated in $\cN=(1,0)$
harmonic superspace \cite{HSW}, \cite{Z}, \cite{BIS} (for the
harmonic superspace approach, see \cite{GIKOS}, \cite{GIOS}). The
theory is quantized in the framework of the harmonic superfield
background method which was originally developed for $4D, \cN=2$ SYM
theories in \cite{BBKO}, \cite{BK}, \cite{BBK}\footnote{Review of
various applications of the background harmonic superfields for
studying the effective actions of $4D, \cN=2,4$ SYM theories was
given in \cite{BIP-rev}, \cite{BISam}.} and then generalized to $6D,
\cN=(1,0)$ gauge theories in \cite{BIM}, \cite{BIMS-a},
\cite{BIMS-b}, \cite{BIMS-c}.

Note that $6D, \cN=(1,1)$ SYM theory is non-renormalizable by power
counting. However, it was found that this theory is on-shell finite
at one and two loops \cite{FT}, \cite{MarSag1}, \cite{MarSag2},
\cite{HS}, \cite{HS1}, \cite{BHS}, \cite{BHS1}, \cite{Bork}.
Recently,  the aspects of renormalizability of the theory under
consideration were studied by harmonic superspace techniques. It was
shown that there is a gauge choice at which the one-loop divergences
are completely canceled off shell \cite{BIMS-a}, \cite{BIMS-b},
\cite{BIMS-c}. Some two-loop harmonic supergraphs are also finite
\cite{BIMS-d}\footnote{Gauge dependence of the one-loop divergences
was studied in \cite{BIMS6}.}. These results guarantee that the
one-loop effective action for the background fields satisfying the
classical equations of motions is finite.

The paper is organized as follows. In Section 2 we recall the
formulation of the six-dimensional $\cN=(1,1)$ SYM theory in terms
of interacting $\cN=(1,0)$  gauge multiplet and hypermultiplet.
Assuming that the hypermultiplet is in the adjoint representation of
gauge group, we gain an additional implicit $\cN=(0,1)$
supersymmetry and, as a result, obtain the complete $\cN=(1,1)$
supersymmetric gauge theory. Then, in Section 3, we formulate the
one-loop effective action of the theory in the framework of the
superfield background method. The effective action constructed in
this way depends on all fields of $\cN=(1,1)$ gauge  multiplet. We
restrict our consideration to the slowly varying background
superfields which are solution of the classical equations of motion,
since such an approximation suffices for finding out the low-energy
effective action. Section 4 is devoted to deriving the complete
one-loop Heisenberg-Euler-type superfield effective action. We
follow the $6D, \cN=(1,0)$ harmonic superspace version of the
procedure developed in \cite{KM2}, \cite{K}, construct the
superfield heat kernel for the Green function of the gauge multiplet
and find the explicit expression for the Green function in the
coincident-points limit. This expression directly yields the
low-energy effective action. Section 5 contains a brief summary of
our results and a list of possible further studies. In Appendix we
give the definition and outline the basic properties of the parallel
displacement operator \cite{K} in the harmonic superspace. This
operator is of need while constructing the heat kernel for the Green
function of gauge multiplet.


\section{The model and conventions}

We formulate the $6D, \,\cN=(1,1)$ SYM theory in terms of the gauge
multiplet $V^{++}$ and hypermultiplet $q^{+}_A$. Both these harmonic
superfields satisfy the Grassmann analyticity conditions
$D^{+}_{a}V^{++}=0$ and $ D^{+}_{a} q^{+}_{A}=0$, where the spinor
derivative in the analytic basis \cite{GIOS} reads $D^+_a =
\tfrac{\partial}{\partial \theta^{-a}}$. The superfield action of
$\cN = (1,1)$ SYM theory is written as a sum of the actions for the
gauge multiplet and for the hypermultiplet
 \bea
 S_0[V^{++}, q^+]&=&
\frac{1}{{\rm f}^2}\Big\{\sum\limits^{\infty}_{n=2}
\frac{(-i)^{n}}{n} \tr \int d^{14}z\, du_1\ldots du_n
\frac{V^{++}(z,u_1 ) \ldots
V^{++}(z,u_n ) }{(u^+_1 u^+_2)\ldots (u^+_n u^+_1 )}  \nn \\
&& - \f12 \tr \int d\zeta^{(-4)} du\, q^{+\,A} \nabla^{++}
q^{+}_{A}\Big\}\,, \label{S0}
 \eea
where ${\rm f}$ is a dimensionful coupling constant ($[{\rm
f}]=-1$). We include the integration over harmonics into the
integration measure over the analytic subspace, $d\zeta^{(-4)} = d^6
x_{(\rm an)}\, du\,(D^-)^4$. In the action \eq{S0} the
hypermultiplet is minimally coupled with the gauge multiplet by
means of the covariant harmonic derivative $\nb^{++}$,
 \be
\nabla^{++}q^{+}_{A} = D^{++} q^{+}_{A} + i [V^{++},q^{+}_{ A}]\,.
\label{Vfirst}
 \ee
The action \eq{S0} is invariant under the infinitesimal gauge
transformations
 \bea
 \d V^{++} = -\nb^{++} \Lambda\,,  \quad \d q^{+}_{ A} =  i[\Lambda, q^{+}_{A}]\,,
 \label{gtr}
 \eea
where $\Lambda(\zeta, u) = \widetilde{\Lambda}(\zeta, u)$ is a real
analytic gauge parameter.

We also introduce the non-analytic superfield $V^{--}$ as a solution
of the zero curvature condition \cite{GIOS}
 \bea
 D^{++} V^{--} - D^{--}V^{++} + i [V^{++},V^{--}]=0\,, \label{zeroc}
 \eea
and define one more covariant harmonic derivative $\nb^{--} = D^{--}
+ i V^{--}$. Using the superfield $V^{--}$ we construct the ${\cal
N}=(1,0)$ gauge superfield strength
 \bea
 W^{+a}=-\f{i}{6}\varepsilon^{abcd}D^+_b D^+_c D^+_d V^{--}
 \eea
with the useful off-shell properties
 \bea
\nabla^{++} W^{+a}  = \nabla^{--} W^{-a} \ =\ 0\,, \qquad W^{-a} =
\nabla^{--}W^{+a} \,. \label{HarmW}
 \eea

Let we introduce an analytic superfield $F^{++}\,$,
 \bea
 F^{++} = (D^+)^4 V^{--}\,, \qquad D^+_a F^{++} = \nb^{++}
 F^{++}=0\,,
 \eea
and evaluate the classical equations of motion corresponding to the
action \eq{S0}
 \bea
 F^{++}  + \tfrac12 [q^{+ A},q^{+}_{ A}] = 0\,, \qquad  \nabla^{++}\,q^{+}_{A} = 0\,.
\label{Eqm}
 \eea

We assume that both $V^{++}$ and $q^{+A}$ superfields take values in
the adjoint representation of the gauge group. Hence the action
\eq{S0} possesses the extra implicit  ${\cal N}=(0,1)$
supersymmetry,
 \be
\delta_0 V^{++} = \epsilon^{+ A}q^+_A\,, \quad \delta_0 q^{+ A} =
-i(D^+)^4 (\epsilon^-_A V^{--})\,, \quad \epsilon^{\pm}_A =
\epsilon_{aA}\theta^{\pm a}\,,\label{Hidden}
 \ee
which completes the manifest ${\cal N}=(1,0)$ supersymmetry to
${\cal N}=(1,1)\,$. It is convenient to use the following
representation for the variation $\delta_0q^{+ A}$
 \be
\label{deltaq} \delta_0 q^+_A =- \epsilon_{aA}(\theta^{-a} F^{++} -
W^{+ a}),
 \ee
which is exspressed through the superfield strengths $F^{++}$ and
$W^+_a$.


\section{One-loop effective action in the  background superfield method}

Let we apply the background superfield method to the six-dimensional
SYM theory\footnote{The background superfield method for $4D, {\cal
N}=2$ gauge theories in harmonic superspace was worked out in
\cite{BBKO} and generalized for six-dimensional gauge theory in
$\cN=(1,0)$ harmonic superspace in the works
\cite{BIMS-a},\cite{BIMS-b},\cite{BIMS-c}.}. Following to the method
we split the superfields $V^{++}, q^{+}$ into the sum of the
``background'' superfields $V^{++}, Q^{+}$ and the ``quantum'' ones
$v^{++}, q^{+}\,$,
 \be
 V^{++}\to {\bf V}^{++} + {\rm f} v^{++}, \qquad q^{+}_A \to {\bf Q}^{+}_A + {\rm
 f}q^{+}_A\,.
 \ee
Then we have to expand the action in a power series with respect to
the quantum fields. The one-loop quantum correction $\Gamma^{(1)}$
to the classical action for the model \eq{S0} is given by
 \be
 e^{ i\Gamma^{(1)}[{\bf V}^{++}, {\bf Q}^+]} =\mbox{Det}^{1/2}\sB \int {\cal
D}v\,{\cal D}q\, {\cal D}{\bf b}\,{\cal D}{\bf c}\,{\cal
D}\varphi\,\,\, e^{iS_{2}[v^{++}, q^+, {\bf b}, {\bf c}, \varphi,
{\bf V}^{++}, {\bf Q}^+]}\,,
 \label{Gamma0}
 \ee
where
  \bea
 S_2 &=& S_{gh}-\frac{1}{2}\tr\int d\zeta^{(-4)}du\, v^{++}\sB v^{++}
  -\f12 \tr\int d\zeta^{(-4)}du\,{q}^{+ A}\nb^{++} q^{+}_{A} \nn \\
 &&- \f{i}{2}\tr \int d\zeta^{(-4)}du\Big\{
  {\bf Q}^{+ A}[ v^{++}, q^{+}_{ A}] + {q}^{+ A}[v^{++}, Q^{+}_{A}]\Big\}\,,
  \label{S2}\\
  S_{gh} &=& \tr \int d\zeta^{(-4)}\,{\bf b}(\nb^{++})^{2}{\bf c}
 + \frac{1}{2}\tr\int
 d\zeta^{(-4)}\,\varphi(\nb^{++})^{2}\varphi\,. \label{Sgh}
 \eea
The action for ghosts superfields $S_{gh}$ \eq{Sgh} involves the
actions for the Faddeev-Popov ghosts ${\bf b}$ and ${\bf c}$ and
also for the Nielsen-Kallosh ghost $\varphi$. The
covariantly-analytic d'Alembertian $\sB$ is defined as
$\sB=\frac{1}{2}(D^+)^4(\nb^{--})^2$, where the harmonic covariant
derivative $\nb^{--} = D^{--} + i {\bf V}^{--}$ contains the
background superfield ${\bf V^{--}}$. While acting on an analytic
superfield, the operator $\sB$ is given by
\begin{eqnarray}
&& \sB= \eta^{MN} \nabla_M \nabla_N + {\bf W}^{+a} \nabla^{-}_a +
{\bf F}^{++} \nabla^{--} - \frac{1}{2}(\nabla^{--} {\bf F}^{++})\,,
\label{Box_First_Part}
\end{eqnarray}
where $\eta^{MN} = {\rm diag} (1, -1, -1, -1, -1, -1)$ is the
six-dimensional Minkowski metric, $M,N=0,..,5$, and
$\nabla_{M}=\pr_M + i{\bf A}_M$ is the background-dependent vector
supercovariant derivative (see \cite{BIS} for details).

The action $S_2$ (\ref{S2}) contains terms with mixed quantum
superfields $v^{++}$ and $q^{+}$. For further use, we diagonalize
this quadratic form by means of the special substitution of the
quantum hypermultiplet variables in the path integral \eq{Gamma0},
such that it removes the mixed terms,
 \bea
 \label{replac}
 q^{+}_{A}(1)= h^{+}_{A}(1) - i\int d \zeta^{(-4)}_2 du_2\, G^{(1,1)}(1|2)_A{}^B
 [v^{++}(2), Q^{+}_{B}(2)]\,,
 \eea
with $h^{+}_{n}$ being a set of new independent quantum superfields.
It is evident that the Jacobian of the variable change
(\ref{replac}) is equal to one. Here
$G^{(1,1)}(\zeta_1,u_1|\zeta_2,u_2)_A{}^B = i\langle0| {\rm
T}{q}^{+}_{A}(\zeta_1,u_1) \tilde{q}^{+\,B}(\zeta_2,u_2)|0\rangle$
is the superfield hypermultiplet Green function in the $\tau$-frame.
This Green function is analytic with respect to its both arguments
and satisfies the equation
 \bea
 \label{eqG}
 \nb_1^{++}G^{(1,1)}(1|2)_A{}^B &=&\delta_A{}^B \d_{\cal A}^{(3,1)}(1|2)\,.
 \eea
In the $\tau$-frame the Green function can be written in the form
$G^{(1,1)}(1|2)_A{}^B = \d_A{}^B G^{(1,1)}(1|2)$, where
 \bea
 G^{(1,1)}(1|2) &=&  \f{(\nb^+_1)^4(\nb^+_2)^4}{\sB_1}\f{\d^{14}(z_1-z_2)}{(u^+_1u^+_2)^3}\,.
 \label{GREEN}
  \eea
Here $\d_A^{(3,1)}(1|2)$ is the covariantly-analytic delta-function.

After performing the shift \eq{replac}, the quadratic part of the
action $S_2$ \eq{S2} splits into few terms, each being bilinear in
quantum superfields:
 \bea
 S_2 &=& S^v_{2}- \tr\int d\zeta^{(-4)}du\, h^{+A}
 \nb^{++}h^{+}_A + \tr \int d\zeta^{(-4)}du\,{\bf b}(\nb^{++})^{2}{\bf c} \nn \\
 &&
  + \frac{1}{2}\tr\int d\zeta^{(-4)}du\,\varphi(\nb^{++})^{2}\varphi \label{S22}\\
 S_2^v &=& \frac{1}{2} \tr\int d\zeta_1^{(-4)}\,d\zeta_2^{(-4)}\,
 du_1 du_2\,
 v_1^{++}\Big\{\sB \d^{(3,1)}_A(1|2)
 - {Q}^{+ A}(1) G^{(1,1)}(1|2)Q^{+}_{A}(2)\Big\}v_2^{++}\,.
 \label{S2v}
 \eea

In the action \eq{Gamma0} the background superfields ${\bf V^{++}} $
and ${\bf Q}^+$ are analytic but unconstrained otherwise. The gauge
group of the theory \eq{S0} is assumed to be $SU(N)$. For the
further consideration, we will also assume that the background
fields ${\bf V}^{++}$ and  ${\bf Q^+}$ align in a fixed direction in
the Cartan subalgebra of $su(N)$
 \bea
 {\bf V}^{++} = V^{++}(\zeta,u) H\,, \qquad {\bf Q^+} =  Q^+(\zeta,u)\, H\,,
 \eea
where $H$ ia a fixed generator in  the Cartan subalgebra generating
some abelian subgroup $U(1)$ \footnote{We denote the $H$ component
of ${\bf V}^{++}$ by the same letter $V^{++}$ as the original
non-abelian harmonic connection, with the hope that this will not
create a misunderstanding. The same concerns the abelian superfield
strength $W^{+ a}$.}. Our choice of the background corresponds to
the spontaneous symmetry breaking $SU(N) \rightarrow SU(N-1)\times
U(1)$.

The classical equations of motions \eq{Eqm} for the background
superfields $V^{++}$ and  $\Omega$ are free
 \bea
 F^{++} = 0\,, \qquad D^{++} Q^+_A = 0\,. \label{EqmB}
 \eea
In that follows we assume that the background superfields solve the
classical equation of motion \eq{EqmB}. We will also assume that the
background is slowly varying in space-time, {\it i.e.},
 \bea
 \partial_M W^{+a} \simeq 0\,, \qquad \partial_M Q^+_A \simeq 0\,.
 \label{constBG}
 \eea

Thus we end up with an abelian background analytic superfields
$V^{++}$ and $Q^+_A$, which satisfy the classical equation of motion
\eq{EqmB} and the conditions \eq{constBG}. Under these assertions
the gauge superfield strength $W^{+a}$ is analytic \footnote{ In
general this is not true and $F^{++}\neq0$.}, $D^+_a W^{+b} = \d_a^b
F^{++}=0$. For further analysis it is convenient to use the
$\cN=(0,1)$ transformation for gauge superfield strength $W^{+a}$
\cite{BIS}. In the case of the slowly varying abelian on-shell
background superfields the hidden $\cN=(0,1)$ supersymmetry
transformations have a simple form,
 \bea
  \delta Q^+_A = \epsilon_{a A} W^{+ a}\, \qquad \delta W^{+ a}=0\,.
  \label{Onshell2}
 \eea
It is worth pointing out that these conditions  are covariant under
${\cal N}=(0,1)$ supersymmetry by themselves.

We  choose the Cartan-Weyl basis for the $SU(N)$ gauge group
generators, so that the quantum superfield $v^{++}$ has the
decomposition
 \bea
 v^{++} =  v^{++}_{\rm i} H_{\rm i}+ v^{++}_\a E_\a\,, \qquad {\rm i} = 1,..,
 N-1,\quad \a = 1,..,N(N-1)\,,
 \eea
where $E_\a$ is the generator corresponding to the root $\a$
normalized as $\tr(E_\a E_{-\b} )=  \d_{\a\b}$ and $H_{\rm i}$ are
the Cartan subalgebra generators, $[H_{\rm i}, E_\a] = \a_{H_i}
E_\a$. In this case the background covariant d'Alembertian
\eqref{Box_First_Part} under the conditions \eq{EqmB} acts on the
quantum superfield $v^{++}$ as
 \bea
 \sB v^{++}&=& \f12 (D^+)^4 \Big\{ (D^{--})^2 v^{++}
 + i \a_H D^{--} V^{--}v_\a^{++} E_\a
 \nn \\ &&  \qquad\qquad + i \a_H V^{--} D^{--}
 v^{++}_\a E_\a  - \a^2_H(V^{--})^2 v^{++}_\a E_\a \Big\} \\
  &=& \sB_H\, v^{++}_\a E_\a + \pr_M\pr^M\, v^{++}_{\rm i} H_{\rm i}
  \,,
 \eea
where we have introduced the operator
 \bea
 \sB_{H} := \nb^{ab}\nb_{ab} + \a_{H}\, W^{+a} D^-_a \,. \label{sboxH}
 \eea

The one-loop effective action \eq{Gamma0} with  the action $S_2$
\eq{S22} for the background superfields $V^{++}$ and $Q^+$ subjected
to the conditions \eq{EqmB} and \eq{constBG} thus reads
 \bea
\Gamma^{(1)}&=& \f{i}2\Tr_{(2,2)} \ln\Big(\sB_H - \alpha_H^2 {Q}^{+
A} G^{(1,1)}Q^{+}_{A}\Big) - \f{i}2\Tr_{(4,0)} \ln \sB_H\,.
\label{1loop}
 \eea
The first term in he expression \eq{1loop} is the contribution from
the gauge multiplet \eq{S22}, while the second one comes from
$\mbox{Det}^{1/2}\sB $ in \eq{Gamma0}. The contributions from the
Faddeev-Popov and Nielsen-Kallosh ghosts are canceled by the
contribution from quantum hypermultiplet.

We use the standard definition for the functional trace over
harmonic superspace in \eq{1loop}
$$\Tr_{(q,4-q)} {\cal O} = \tr \int d \zeta_1^{(-4)}d \zeta_2^{(-4)}
\, \d_{\cal A}^{(q,4-q)}(1|2)\,  {\cal O}^{(q,4-q)}(1|2)\,.
$$
Here $ \d_{\cal A}^{(q,4-q)}(1|2)$ is an  analytic delta-function
\cite{GIOS} and  ${\cal O}^{(q,4-q)}(\zeta_1,u_1|\zeta_2,u_2)$ is
the  kernel of an operator acting in the space of analytic
superfields with the harmonic U(1) charge $q$.

As the next step, we rewrite the contribution from
$\mbox{Det}^{1/2}\sB $ as the functional integral over a zero-charge
analytic superfield $\sigma$ with the action
 \bea
-\f12\tr \int d\zeta^{(-4)} \sigma (\nb_H^{++})^2 \sB_H \sigma \,,
\label{det}
 \eea
where $\nabla_H^{++} = D^{++} + i \alpha_H V^{++}$. Then we divide
the superfield $v^{++}$ into the two orthogonal projections (see the
reviews \cite{BIP-rev},\cite{BISam})
 \bea
  v^{++} = v^{++}_{\rm T} + \nb^{++}_H \xi\,, \qquad \nb^{++}_H
  v^{++}_{\rm T} = 0\,. \label{subs}
 \eea
The transversal component $v^{++}_{\rm T}$ of the superfield
$v^{++}$ is defined as
 \bea
 v^{++}_{\rm T}(1) = \int d\zeta^{(-4)}_2 du_2 \Pi_{\rm T}^{(2,2)}(1|2)
 v^{++}_2\,,
 \eea
where $\Pi^{(2,2)}_{\rm T}(\zeta_1,u_1; \zeta_2,u_2)$ is the
projector on the space of covariantly analytic transverse
superfields. After substitution of $v^{++}$ \eq{subs} in the
quadratic part of the action for the gauge multiplet \eq{S2v} we
obtain an additional contribution from the bosonic superfield
$\xi\,$,
 \bea
\f12 \tr \int d\zeta^{(-4)}du\, \xi (\nb^{++}_H)^2 \sB_H \xi.
 \eea
Note that all mixed terms vanish due to the properties $\nb^{++}_H
v^{++}_{\rm T} = 0$ and $Q^{+A} Q^+_A = 0$.

The contribution from the superfields $\xi$ and $\sigma$ cancel each
other in the one-loop effective action and finally we obtain
 \bea
 \Gamma^{(1)}= \f{i}2 \Tr_{\rm T} \ln \Big(\sB_H - \alpha_H^2 {Q}^{+
A} G^{(1,1)}Q^{+}_{A}\Big)\,, \label{1l}
 \eea
where trace is over the space of analytic superfields $v^{++}_{\rm
T}$ constrained by the condition $\nb^{++}_H v^{++}_{\rm T}~=~0$.

Let us consider the quadratic action which produces the effective
action \eq{1l},
  \bea
  S^{(2)} = \f12\tr\int d\zeta^{(-4)}_1 d \zeta^{(-4)}_2\, du_1 du_2\,
 v_{\rm T}^{++}(1)\Big\{\sB_H \d^{(3,1)}_A(1|2)
 - \a_H^2{Q}^{+ A}(1) G^{(1,1)}(1|2)Q^{+}_{A}(2)\Big\}v_{\rm T}^{++}(2)\,. \nn
   \eea
First of all we study the non-local term $Q^{+A}(1)
G^{(1,1)}(1|2)Q^{+}_A(2)$ in this expression in the  coincident
harmonic points ($u_2\to u_1$) limit. We rewrite the Green function
$G^{(1,1)}$ as follows \cite{Kuz01}
\begin{eqnarray}
G^{(1,1)}(1|2) &=&\frac{(D_1^+)^4}{\sB}\Big\{ (D_1^-)^4 (u^+_1u^+_2)
- \Omega^{--}_1(u^-_1 u^+_2) + \sB\frac{(u^-_1 u^+_2)^2}{(u^+_1
u^+_2)} \Big\} \delta^{14}(z_1-z_2), \label{OmegaG}
\end{eqnarray}
where $ \Omega^{--} = i\nb^{ab}\nb^-_a \nb^-_b - W^{-a}\nb^-_a +
\frac{1}{4} (\nb^-_a W^{- a})$. According to its definition
\eq{GREEN}, the Green function $G^{(1,1)}(1|2)$
 is analytic with respect to its  both
arguments. The representation \eq{OmegaG} preserves the analyticity
in the second argument, though  in some implicit way (see, e.g.,
\cite{Kuz01}).

The third term in \eq{OmegaG} is singular in the $u_2\to u_1$ limit.
To avoid the singularity, we expand  $Q^+_{A}(2)$ over harmonics
using the property $Q^+_A(2) = (u^+_1 u^+_2)Q_A^-(1) - (u^-_1
u^+_2)Q_A^+(1)$ \cite{Kuz01} and reconstruct the full integration
measure by taking off the $(D^+_1)^4$ factor from Green function in
\eq{OmegaG}. We obtain the non-singular expression
 \bea
 -\f{\a^2_H}2 \tr \int d^{14}z_1 d \zeta^{(-4)}_2\, du_1 du_2\,
 v_{\rm T}^{++}(1) v_{\rm T}^{++}(2)  Q^{+A}Q^{-}_A(1) (u^-_1 u^+_2)^2
 \delta^{14}(z_1-z_2) + \ldots, \label{nonsing}
 \eea
where dots stand for the rest of terms coming from the expansion of
Green function $G^{(1,1)}(1|2)$ in \eq{OmegaG}. These terms are
proportional to $(u_1^+u_2^+)$ and vanish in the effective action
for the on-shell background due to property $(u^+_1 u^+_2)|_{2\to 1}
=0$ \cite{GIOS}.

The combination $Q^{+A}Q^-_A$ is a gauge invariant real superfield.
However, the superfield $Q^{+A}Q^-_A$ is not analytic and only the
full expression \eq{nonsing} preserves the analyticity. For further
consideration it will be convenient to replace the background
hypermultiplet $Q^{+}_A$ by the analytic omega-hypermultiplet
$\Omega$, using the correspondence \cite{GIOS}
 \bea
 Q^+_A = u^+_A \Omega - u^-_A D^{++} \Omega\,.
 \label{change}
 \eea
The $\cN=(0,1)$ supersymmetry transformation of $Q^{+}_A$ defined in
\eqref{Onshell2} implies the following transformation law for the
superfield $\Omega$:
 \bea
 \d \Omega = \epsilon^-_a W^{+ a}\,, \qquad \delta (D^{++}\Omega)
 = \epsilon^+_a W^{+ a}\,, \qquad \d W^{+a} = 0\,.\label{Onshell3}
 \eea
The on-shell condition \eq{EqmB} for $Q^+_A$ and the  definition
\eq{change}  give rise to the equation of motion for the $\Omega$
hypermultiplet in the form
 \bea
 (D^{++})^2 \Omega = 0\,. \label{Onshell4}
 \eea

Now let us discuss the possible structure of effective action
\eq{1l} after passing from the background  $Q^{+}_{A}$
hypermultiplet to the $\Omega$ hypermultiplet  by eq. \eq{change}.
We assume that the background superfields satisfy the classical
equations of motion and slowly vary in space-time.  As shown in
\cite{BIM}, the hidden $\cN=(0,1)$ supersymmetry severely restricts
the possible structure of the effective action \eq{1l}. We consider
the analytic contributions to the effective action which respect the
implicit $\cN=(0,1)$ supersymmetry \eq{Onshell3} and are local in
harmonic superspace. Thus for $\Gamma^{(1)}$ we should have the
following general expression:
 \bea
 \Gamma^{(1)}=\int d\zeta^{(-4)} du\, (W^+)^4 {\cal F}(\Omega, D^-_a W^{+b})\,,
 \label{G_gen}
 \eea
where ${\cal F}(\Omega, D^-_a W^{+b})$ is a real analytic function
with zero harmonic $U(1)$ charge. Here we have to emphasize that
within our approximation the function ${\cal F}$ can depend only on
the background superfield $\Omega$ and $D^-_a W^{+b}$. Indeed,
including, e.g., the contributions with harmonic derivative $D^{++}$
of superfield $\Omega$ will amount to the necessity  to compensate
the extra harmonic charge $+2$. One can accomplish this, acting on $
D^{++}\Omega$ by the spinor derivatives with negative charge, {\it
i.e.} by passing to $D^-_a D^-_b D^{++}\Omega$. Moreover, such
contributions are analytic in the constant background approximation
which we use. But the covariant d'Alembertian \eq{sboxH} includes
the operator $D^-_a$ multiplied by the background superfield
strength $W^{+a}$. Thus all contributions of the kind $D^-_a D^-_b
D^{++}\Omega$ have to contain $W^{+a}$ and so they immediately
vanish due to the presence of the maximal power of $(W^{+})^4$ in
the integrand of \eq{G_gen}. Also we exclude from the consideration
all contributions containing $D^{--}D^{++}\Omega$. Such terms are
not analytical and do not contribute to the effective action. So in
that follows we take into account only the contributions having no
harmonic derivatives of the background superfield $\Omega$.

Keeping in mind this discussion, we rewrite the one-loop effective
action \eq{1l}, applying the proper-time method
 \bea
 \Gamma^{(1)} = -\frac{i}2 \tr \int d \zeta_1^{(-4)} du_1\, \int_0^{\infty}
 \frac{d s}{s} e^{is(\sB_1 - \alpha_H^2 \Omega^2 )}\Pi_{\rm T}^{(2,2)}(1|2)\bigg|_{2=1}\,.
 \eea
The covariant analytic projector $\Pi_{\rm T}^{(2,2)}(1|2)$ in the
limit $u_2 \to u_1$ has the simple form \cite{BIM,Kuz01}
 \bea
\Pi_{\rm T}^{(2,2)}(1|2)\bigg|_{u_2=u_1} =
-(D_1^{+})^4\delta^{14}(z_1-z_2)\,.
 \eea
Also, in order  to avoid the dependence of the effective action on
the root $\alpha_H$,  we have to calculate the trace over matrix
indices. We will consider the simplest case, when the gauge group of
the theory is $SU(2)$.  We obtain the following final expression for
the one-loop effective action
 \bea
\Gamma^{(1)} = i\int d \zeta_1^{(-4)} du_1\, \int_0^{\infty}
 \frac{d s}{s} e^{is(\sB_1 - \Omega^2 )}(D_1^{+})^4\delta^{14}(z_1-z_2)\bigg|_{2=1}\,.
 \label{1loop3}
  \eea
The expression \eq{1loop3} is the central object of our further
consideration. In the next section we will calculate it under the
simplifying assumptions on the background superfields formulated
earlier.

\section{Complete contribution to one-loop effective action }

To find the complete low-energy effective Lagrangian we should
calculate \eq{1loop3}. We use the covariantly constant on-shell
gauge and omega-hypermultiplet background superfields subject to the
constraints \eq{EqmB} and \eq{Onshell4}. We also introduce the
notation
 \bea
 D^-_a W^{+b} = - D^+_a W^{-b} = N_a^b\,, \label{backgr}
 \eea
where the superfield $N_a^b$ is related to the gauge field strength
$F_a^b = i (\sigma^{MN})_a^b F_{MN}$ as
 \bea
F_a^b =  D^-_a W^{+b} - D^+_a W^{-b} = 2N_a^b\,.
 \eea
We use the following definition for the generator of spinor
representation $(\sigma^{MN})_a^b$
 \bea
 (\sigma_{MN})^a{}_b = \f12 (\tilde{\gamma}^M \gamma^N - \tilde{\gamma}^N
 \gamma^M)^a{}_b\,,
 \eea
where the antisymmetric six-dimensional $(\gamma^M)_{ab}$ and
$(\tilde{\gamma}^M)^{ab}$ matrices are related as
 \bea
 (\tilde{\gamma}_M)^{ab} = \tfrac12\varepsilon^{abcd}(\gamma_M)_{cd}\,,
 \eea
and $\varepsilon^{abcd}$ is the totally skew-symmetric $6D$ tensor.
The matrices $\gamma_M$ and $\tilde{\gamma}_M$ are subject to the
basic relations for Weyl matrices
 \bea
 (\gamma_M)_{ac} (\tilde{\gamma}_N)^{cb}+ (\gamma_N)_{ac} (\tilde{\gamma}_M)^{cb} = -2 \delta_a{}^b \eta_{MN},
 \qquad (\gamma^M)_{ac} ({\gamma}_M)_{cb} = 2 \varepsilon_{abcd}\,.
  \eea
As before, we choose the Minkowski metric $\eta_{MN}, \, M,N =
0,..,5,$ with the mostly negative signature (see its definition
after eq. (3.5)).

Then, as in the $4D$, $\cN=2$  case \cite{Kuz01}, we introduce the
operator $\Delta$,
 \be\label{sDelta}
 \Delta = \sB - W^{-\a}D^+_\a\,,
 \ee
which coincides with $\sB = \nb^{ab}\nb_{ab} + W^{+a}D^-_a$ on the
space of covariantly analytic superfields\footnote{Note that in
$6D,\, \cN=(1,0)$ hypermultiplet theory, the operator (\ref{sDelta})
differs from the analogical operator in $4D,{\cal N}~=~2$
hypermultiplet theory \cite{Kuz01}.}. Thus the expression
\eq{1loop3} takes the form
 \bea
\Gamma^{(1)}=i\int d \zeta_1^{(-4)} du_1\, \int_0^{\infty}
 \frac{d s}{s} e^{is(\Delta_1 - \Omega^2 )}(D_1^{+})^4\delta^{14}(z_1-z_2)\bigg|_{2=1}\,.
 \eea
Note that the spinor derivative $D^-_a$ can act on the superfield
$W^{-a}$ in the operator $\Delta$. However, the operator $\Delta -
\Omega^2$ standing in the exponential  does not commute with
$(D^+)^4$ even in the case of constant on-shell background. Thus,
pulling the exponential with the argument $\Delta-\Omega^2$ through
$(D^{+})^4$, we obtain
  \bea
\Gamma^{(1)}=i\int d \zeta_1^{(-4)} du_1\, \int_0^{\infty}
 \frac{d s}{s} (e^{-isN}D_1^+)^4 e^{is(\Delta_1 - \Omega^2 )}\delta^{14}(z_1-z_2)\bigg|_{2=1}\,.
  \label{1loop4}
  \eea

Let us introduce the heat kernel for the operator $\Delta -
\Omega^2$,
  \bea
 K(z_1,z_2|s) = e^{is(\Delta_1-\Omega^2)}\delta^{14}(z_1-z_2)\,,
 \eea
as a formal solution of the equation
 \bea
 \Big(i\f{d}{ds} + \Delta_1 - \Omega^2 \Big) K(z_1,z_2|s) =
 \delta^{14}(z_1-z_2)\,.
 \eea
In terms of the kernel $K(z_1,z_2|s)$ the one-loop effective action
\eq{1loop4} can be rewritten as
 \bea
\Gamma^{(1)}=i\int d \zeta_1^{(-4)} du_1\, \int_0^{\infty}
 \frac{d s}{s} (e^{-isN}D_1^{+})^4 K(z_1,z_2|s)\big|_{2=1}\,.
 \label{1loop5}
 \eea

We denote by $\Upsilon$ the first-order operator appearing in
$\Delta$, {\it i.e.} write the latter as follows
 \be
 \label{Upsilon}
\Delta  = \nb^{ab}\nb_{ab}  + \Upsilon\,,  \quad \Upsilon :=
W^{+a}D^-_a - W^{-a}D^+_a\,.
 \ee
We provide the calculation in the case of a covariantly constant
vector multiplet (\ref{constBG}). The vector covariant derivative
$\nb_{ab}$  turns out to commute with the operator $\Upsilon$,  as
well as with the additional term $\Omega^2$. This allows us to
represent $e^{is(\Delta-\Omega^2)}$ in the factorized form
$e^{is(\Upsilon-\Omega^2 )}e^{is\nb^{ab}\nb_{ab}}$ and to calculate
the heat kernel $K(z_1,z_2|s)$
 \be
 K(z_1,z_2|s)=e^{is(\Upsilon-\Omega^2)}e^{is\nb^{ab}\nb_{ab}}\d^{14}(z_1-z_2)
 = e^{is(\Upsilon- \Omega^2)}\tilde{K}(z_1,z_2|s).
 \ee
The further steps in calculation of \eq{1loop5} are similar to those
performed in \cite{BMP}. We use the momentum representation of the
delta function, $\delta^{14}(z_1-z_2)=\delta^6(x_1-x_2)
\delta^4(\theta^{+}_1-\theta^{+}_2)\delta^4(\theta^{-}_1-\theta^{-}_2)$,
 \be\label{delta}
 {\bf 1}\d^{(14)}(z_1-z_2)=\int \f{d^6 p}{(2\pi)^6}
 e^{i\rho^M p_M} \zeta^{+4} \zeta^{-4} I(z_1,z_2)~,
 \ee
where $I(z_1,z_2)$ is a parallel displacement operator in superspace
\cite{KM2,K} (see details in Appendix) and
 \bea
 \rho^M=(x_1-x_2)^M -2i\zeta^{+ a}(\g^M)_{ab}\t^{- b}_1, \q
 \zeta^{\pm a}=(\t^\pm_1-\t^\pm_2)^a~.
 \label{zeta}
 \eea

The reduced heat kernel $\tilde{K}(z_1,z_2|s)$ can now be evaluated
in the same way by generalizing the Schwinger construction
\cite{KM2},
 \bea\label{sch}
 \tilde{K}(z_1,z_2|s)=\f{i}{(4\pi i s)^3}\det{^{\f12}}\bigg(\f{sF}{\sinh sF}\bigg)
 e^{\f{i}{4}\rho^M(F\coth s F)_{MN}\rho^N} \zeta^{+4} \zeta^{-4} I(z_1,z_2)\,,
 \eea
where the determinant is taken with respect to Lorentz indices. To
compute the kernel $K(z_1,z_2|s)$ we need to evaluate the action of
$e^{is\Upsilon}$ on  $\tilde{K}(z_1,z_2|s)$. However, the operator
$\Upsilon$ does not commute with $\Omega^2$ even on shell. To
separate its contribution in $\exp{(is(\Upsilon-\Omega^2))}$,  we
use the Baker-Campbell-Haussdorf formula
 \bea
 e^{is(\Upsilon - \Omega^2)} = e^{(-is\Omega^2 +
 \f{(is)^2}{2}[\Upsilon,\Omega^2]
 -\f{(is)^3}{3!}[\Upsilon,[\Upsilon,\Omega^2]]+\dots)}\,\, e^{is \Upsilon} . \label{BKH1}
 \eea
Using the explicit expression for the commutator $[\Upsilon,
\Omega^2] = W^{+a}(D^-_a\Omega^2) $, one can show that the series in
eq. \eq{BKH1} can be summed up to  the concise expression
 \bea
  e^{is(\Upsilon - \Omega^2)}  = e^{
\big(\f{\exp(-is \,W^{+}D^-)\,-1}{W^{+}D^-}
  \big)\Omega^2} e^{is \Upsilon}. \label{BKH2}
 \eea
The complete structure of the last expression is rather complicated
but it does not matter.  It is crucial for us that it has the form
 \bea
e^{ \big(\f{\exp(-is \,W^{+}D^-)\,-1}{W^{+}D^-} \big)\Omega^2} =
e^{-is \Omega^2 + W^{+a}f_a(W^+,\,N,\,\Omega^2,\,s)}\,,
 \eea
where the function $f_a(W^+,\,N,\,\Omega^2,\,s)$ encodes the whole
information about the series \eq{BKH2}.

As the next step, we act by the operator $e^{is\Upsilon}$ on the
kernel $\tilde{K}(z_1,z_2|s)$. The formal result reads
 \bea
 \label{ker}
 K(z_1,z_2|s) = \f{i}{(4\pi is)^3}\det{}^{\f12}\bigg(\f{sF}{\sinh sF}\bigg)
e^{\f{i}{4}\rho^M(s)(F\coth s F)_{MN}\rho^N(s)} \zeta^{+4}(s)
\zeta^{-4}(s)I(z_1,z_2|s)\,,
 \eea
where we denoted,
 \bea
\zeta^A(s)=e^{is\U}     \zeta^A e^{-is\U}\,,  \quad I(z_1,z_2|s) =
e^{is\U} I(z_1,z_2)\,,
 \eea
and $\zeta^A=(\r^a, \zeta^{\pm a})$.  Using the formula $e^A B
e^{-A} = B+[A,B]+\ldots$ and our constraints on the background
\eq{backgr}, we obtain\footnote{Here we use $D^+_a \zeta^{b-}
=\d_a^b$ and $D^-_a \zeta^{b+} =-\d_a^b$.}
 \bea
 \zeta^{+a}(s) &=&\zeta^{+a} - W^{+b}{\cal N}_b^a \,, \qquad
 \qquad\qquad
 \zeta^{-a}(s) = \zeta^{-a} - W^{-b}{\cal N}_b^a \,, \label{rho} \\
 \r^M(s) &=& \r^M - 2 \int_0^s dt\, W^{-a}(t) (\g^M)_{ab} \zeta^{+b}(t)\,, \qquad
 W^{-a}(s) =  W^{-b}\,\big( e^{isN}\big)^a_b\,.
 \eea
Here we made use of the definition \eqref{backgr} and ${\cal N}_a^b
:= \big(\f{e^{i s N}-1}{N} \big)_a^b$. We do not need the explicit
expression for $I(z_1,z_2|s)$. However,  it is easy to check, by
differentiating with respect to the proper time $s$, that the
following identity holds
 \bea I(z_1,z_2|\,s)
 &=& \exp\left[\int_0^s dt\, \Sigma(z_1,z_2|\,t) \right] I(z_1,z_2)\,,
 \label{I(s)} \\
 \Sigma(z_1,z_2|\,t)&=& e^{it\U} \Sigma(z_1,z_2) e^{-it\U}\,,
  \eea
where $\Sigma(z_1,z_2)$ is defined by the relation
 \be
  (W^{+\a}D^-_a  - W^{-\a}D^+_a)I(z_1,z_2)=\Sigma(z_1,z_2) I(z_1,z_2)\,.
  \label{sigma}
  \ee
For what follows it is important that $\Sigma(z_1,z_2|s)=
W^{+a}\rho_{ab} W^{-b}+\ldots$ (see \eq{sigma2} in the Appendix).

Now we can come back to the calculation of the effective action
\eq{1loop5}. We need to calculate the coincident-points limit for $
(e^{-isN}D_1^{+})^4 K^{z_1,z_2|s}$. The operator
$(e^{-isN}D_1^{+})^4$ acts on the two-point function $\zeta^{-
4}(s)$ and in the coincident-points limit gives the unity
 \bea
(e^{-isN}D^+_1)^4  \zeta^{-4}(s) \Big|_{2=1}&=& 1.
 \eea
For $\zeta^{+4}(s)$ we have
 \bea
\zeta^{+4}(s)\Big|_{2=1} = (W^{+})^4 \det \bigg(\f{e^{i s N}-1}{N}
\bigg)\,. \label{zeta3}
 \eea
We observe that in the coincident-points limit all terms with
$\rho^M(s)$ and $I(z_1,z_2|s)$ have the formal structure
$\exp(W^{+a} +\ldots)$. Due to the presence of the maximal power of
the gauge superfield strength $(W^{+})^4$ in \eq{zeta3} we can
replace the exponential in such terms just by unity.

As the result, we obtain
 \bea \label{G2}
 \Gamma^{(1)} &=& \f{1}{(4\pi)^3}\int d \zeta^{(-4)}du \,(W^{+})^4\,
 \xi\Big(F,N,\Omega^2\Big)\,, \\
 \xi(F,N,\Omega^2) &=& \int_0^{\infty}\f{d s}{s^{4}}\,e^{-s \Omega^2 }\,
\det \bigg(\f{e^{s N}-1}{N}\bigg)
\det{}^{\f12}\bigg(\f{sF}{\sin{sF}}\bigg)\,.
 \label{xi}
 \eea
This is the final expression for the complete low-energy effective
action in the theory under consideration. The effective action
(\ref{G2}), (\ref{xi}) is manifestly gauge invariant and manifestly
$\cN=(1,0)$ supersymmetric by construction. The action \eq{G2} is
also invariant under the implicit $\cN=(0,1)$ supersymmetry
\eq{Onshell3}. Indeed, according to \eq{Onshell3}, the
transformation of $\Omega^2$ is proportional to the superfield
strength $W^{+a}$, $\delta \Omega\sim W^{+a}$. Consequently, all
such terms vanish due to the presence of the maximal power of the
spinorial superfield $(W^+)^4$ in the integrant of \eq{G2}.

In our previous work \cite{BIM} we calculated the leading low-energy
contribution to the one-loop effective action. It has the form
 \bea
 \Gamma^{(1)}_{\rm lead} = \f{1}{(4\pi)^3}\int d\zeta^{(-4)}
 \f{(W^{+})^4}{\Omega^2}\,. \label{G_lead}
 \eea
The expression $\eq{G_lead}$ was obtained under the assumption of
the simplest background, $D^-_a W^{+b} = N_a^b=0$. We see that the
leading contribution \eq{G_lead} immediately follows from \eq{G2}
when $N=F=0$. In this case, $\xi(0,0,\Omega^2) = \f{1}{\Omega^2}$.


\section{Conclusions}

In this paper we considered the quantum aspects of the
six-dimensional $\cN=(1,1)$ SYM theory. We used the $\cN=(1,0)$
harmonic superspace formulation of the theory in terms of
$\cN=(1,0)$ analytic vector gauge multiplet and hypermultiplet. We
assumed that both gauge and matter $\cN=(1,0)$ supermultiplets are
in the adjoint representation of gauge group. By construction, the
theory is invariant under the manifest $\cN=(1,0)$ supersymmetry and
the second implicit $\cN=(0,1)$ one.

We calculated the complete one-loop effective action for the
considered theory in the framework of the background superfield
method in $\cN=(1,0)$ harmonic superspace. We restricted our
attention to the special case of the slowly varying background
superfields satisfying the free classical equations of motion. We
also assumed that background superfields align in the Cartan
subalgebra of $su(2)$. The obtained result \eq{G2} for the effective
action is the complete one-loop effective action for the
six-dimensional $\cN=(1,1)$ SYM theory in the constant background
approximation.

A few comments on the calculation procedure are needed. In six
dimensions the gauge superfield strength is the spinor superfield
$W^{+a}$. The general analysis of the structure of the leading
low-energy effective action \cite{BIM} implies that the effective
Lagrangian as a function of $W^{+a}$ and the $\Omega$ hypermultiplet
has to be an analytic superfield of the $U(1)$ harmonic charge $+4$.
Namely, ${\cal L}^{(+4)} = (W^+)^4 \xi\Big(F,N,\Omega^2\Big)$, where
the function ${\cal \xi}$ was defined in \eq{xi}. It is analytic and
contains the whole information about one-loop quantum corrections.
We have also to recall that, initially, we formulated the theory in
terms of the gauge ${\cal N}=(1,0)$ multiplet and the charge +1
$q^+_A$-hypermultiplet. But during the calculation we were forced to
pass from  the background $q^+_A$-hypermultiplet to the zero-charge
$\Omega$ hypermultiplet. It is known that the matter sector of the
supersymmetric gauge theories can be equivalently described either
by a complex $q^+_A$-hypermultiplet or by a real $\Omega$
hypermultiplet \cite{GIOS}. The reason for making use of $\Omega$ is
that it provides a possibility to define the uncharged analytic
superfield combination playing the role of the background UV cutoff
term in the function $\xi$ \eq{xi}. One can see that the use of the
$q^+_A$-hypermultiplet does not ensure the analyticity required,
because the uncharged combination $q^{+A}q^-_A$ is not analytic.

As the final remark, we emphasize  that there are two interesting
further directions of applying the background field method used
here. One such direction amounts to studying the structure of the
effective action in six-dimensional $\cN=(1,0)$ SYM theory with
higher derivatives \cite{ISZ},\cite{IS},\cite{CT}, another one
concerns deriving the Born-Infeld-type effective action associated
with D5-brane. The latter problem will require carrying out the
superspace multi-loop calculations (see the relevant discussion in
\cite{BPT} on the Born-Infeld-type action related to D3-brane in the
framework of $4D, \cN=4$ SYM theory). Some aspects of the superfield
two-loop calculations of the effective action in $4D, \cN=4$ SYM
theory have been considered in \cite{KM2},
\cite{BPT},\cite{KUZEN1},\cite{KUZEN2},\cite{KUZEN3}.

\section*{Acknowledgments}

\noindent This research was supported in part by RFBR grant, project
No. 18-02-01046, and Russian Ministry of Education and Science,
project No. 3.1386.2017. I.L.B. and B.S.M. are grateful to RFBR
grant, project No. 18-02-00153, for partial support. The work of
B.S.M. was supported in part by the Russian Federation President
grant, the project MK-1649.2019.2.

\section*{Appendix}
\appendix

\section{Parallel displacement operator}
Let us briefly discuss the basic properties of the parallel
displacement operator  $I(z,z')$. By definition, it is defined as a
two-point superspace function depending on the gauge superfields
with the following properties \cite{KM2,K}:
\begin{itemize}
\item[(i)]
Under the gauge transformations it transforms as
 \be
 I(z,z')= e^{i\tau(z)}I(z,z')e^{-i\tau(z')}\,; \label{prop1}
 \ee
\item[(ii)]
It obeys the equation
 \be
\zeta^A \nabla_A I(z,z')=0\,, \label{prop2}
 \ee
where $\zeta^A=(\rho^M, \rho^{a\pm})$ was defined in \eq{zeta};
\item[(iii)] For the coincident
superspace points $z=z'$ it reduces to the identity operator in the
gauge group,
 \be
 I(z,z)=1\,. \label{prop3}
 \ee
\end{itemize}

The general form of the superalgebra of covariant derivatives is as
follows
 \bea
 [\nabla_A,\nabla_B\}={\bf T}_{A\,B}{}^{C}\nabla_C+i{\bf F}_{A\,B}\,,
 \label{supal}
 \eea
where ${\bf T}_{A\,B}{}^{C}$ is a supertorsion and ${\bf F}_{A\,B}$
is a supercurvature for gauge superfield connections. In \cite{KM2}
it was proved that, owing to (\ref{prop2}), the action of the
derivative $\nabla_B$ on $I(z,z')$ can be expressed in terms ${\bf
T}_{AB}{}^C$, ${\bf F}_{AB}$ and their covariant derivatives,
 \bea
 \label{dec2} \nabla_B I(z,z')&=&i
 \sum_{n=1}^\infty \f{(-1)^n}{(n+1)!}\,\bigg[-\zeta^{A_n}
  \ldots\zeta^{A_1}\nabla_{A_1}\ldots\nabla_{A_{n-1}}{\bf F}_{A_n\,B}(z)\\
&& +\, \f{(n-1)}{2}\zeta^{A_n}{\bf T}_{A_n\,B}{}^{C} \zeta^{A_{n-1}}
 \ldots\zeta^{A_1}\nabla_{A_1}\ldots\nabla_{A_{n-2}}{\bf F}_{A_{n-1}\,C}(z)
 \bigg]I(z,z')\,.\nn
 \eea

In our case we do not need the detailed analysis of \eq{dec2}, and
we consider only the simplest background, $N_a^b = 0$. We have
 \bea
D^\pm_a I(z,z') &=& \bigg[\f12 \r_{ab} {W}^{b\, \pm}
-\f{i}{6}(\gamma^M)_{ab}\zeta^{\pm b} \Big( \zeta^{+c} (\gamma_M)_{cd} W^{-d}\nn \\
 &&
  \qquad\qquad \qquad \qquad \qquad +\, \zeta^{-c} (\gamma_M)_{cd} W^{+d}  - i \rho^N F_{NM}\Big)\bigg] I(z,z')\,.
 \eea
Then the superfield $\Sigma(z,z')$ introduced in \eq{sigma}  has the
form
 \bea
\Sigma(z,z') &=& {W}^{+a}\rho_{ab}{W}^{-b}
-\f{i}{6}({W}^{+a}(\gamma^M)_{ab}\zeta^{-b} - {W}^{-a}(\gamma^M)_{ab}\zeta^{+b})\nn \\
 &&\qquad \qquad \times(\zeta^{+c} (\g_M)_{cd}
 {W}^{-d} + \zeta^{-c} (\gamma_M)_{cd} {W}^{+d}  - i \r^N F_{NM})\,.
 \label{sigma2}
 \eea
Thus the decomposition of the superfield $\Sigma(z,z')$  begins with
the gauge superfield strength $W^{+a}$. This is one of the crucial
properties used  in the computation of the coincident-points limit
of the kernel $K(z_1,z_2|s)$.



\end{document}